\documentclass[conference]{IEEEtran}
\IEEEoverridecommandlockouts
\usepackage{comment}
\usepackage{cite}
\usepackage{amsmath,amssymb,amsfonts}
\usepackage{algorithmic}
\usepackage{graphicx}
\usepackage{textcomp}
\usepackage{xcolor}
\usepackage{subfig}
\usepackage{paralist} 
\usepackage{enumitem}
\usepackage{csquotes}
\usepackage{stfloats}
\usepackage{fancyhdr}
\fancypagestyle{firstpage}{%
  \lhead{Accepted for presentation in the Demo Session at the IEEE International Conference on Network Protocols (ICNP), 2024.}}

\usepackage[capitalize,noabbrev]{cleveref}

\def\BibTeX{{\rm B\kern-.05em{\sc i\kern-.025em b}\kern-.08em
    T\kern-.1667em\lower.7ex\hbox{E}\kern-.125emX}}

\usepackage{tikz}
\newcommand*\circled[1]{\:\tikz[baseline=(char.base)]{
            \node[shape=circle,thick,draw,inner sep=1pt] (char) {\footnotesize #1};}}

\setlist{leftmargin=3.5mm}

\usepackage[textsize=tiny, obeyFinal]{todonotes} 

\IEEEoverridecommandlockouts\IEEEpubid{\makebox[\columnwidth]{ 979-8-3503-5171-2/24/\$31.00 $\copyright$2024 IEEE \hfill}\hspace{\columnsep}\makebox[\columnwidth]{ }}
\begin{document}

\title{Demo: Testing AI-driven MAC Learning in Autonomic Networks\\
\thanks{The authors acknowledge the financial support by the German Federal
Ministry of Education and Research (BMBF) in the project “Open6GHub” (grant number 16KISK010, 16KISK011, 16KISK012).}%
}

\author{
\IEEEauthorblockN{
Leonard Paeleke\IEEEauthorrefmark{1}, 
Navid Keshtiarast\IEEEauthorrefmark{2}, 
Paul Seehofer\IEEEauthorrefmark{3},
Roland Bless\IEEEauthorrefmark{3}, \\
Holger Karl\IEEEauthorrefmark{1},
Marina Petrova\IEEEauthorrefmark{2}, and
Martina Zitterbart\IEEEauthorrefmark{3}
}
\IEEEauthorblockA{
\IEEEauthorrefmark{1}\textit{Hasso-Plattner-Institute (HPI)} and \textit{University of Potsdam (UP)} -- firstname.lastname@hpi.de \\
\IEEEauthorrefmark{2}\textit{MCC, RWTH Aachen University} -- \{navid.keshtiarast, petrova\}@mcc.rwth-aachen.de \\
\IEEEauthorrefmark{3}\textit{Karlsruhe Institute of Technology (KIT)} -- firstname.lastname@kit.edu}
}
\maketitle
\thispagestyle{firstpage}
\begin{abstract}
6G networks 
will be highly dynamic, re-configurable, and resilient. 
To enable and support such features, employing AI has been suggested. Integrating AI in networks will likely require distributed AI deployments with resilient connectivity, e.g., for communication between RL agents and environment. Such approaches need to be validated in realistic network environments.
In this demo, we use ContainerNet to emulate AI-capable and autonomic networks that employ the routing protocol KIRA to provide resilient connectivity and service discovery. As an example AI application, we train and infer deep RL~agents learning medium access control (MAC) policies for a wireless network environment in the emulated network.

\end{abstract}



\section{Introduction}

6G networks and beyond
will be highly dynamic due to the inclusion of non-terrestrial networks (e.g., drones, satellites) and increased network virtualization \cite{Organic6G}, among others.
Frequently changing network environments (e.g., topologies, traffic loads) require frequent network re-configurations.
AI has been suggested to optimize network configuration automatically. Proposed use cases range through the entire network stack from network management to optimizing and learning wireless medium access control (MAC) protocols~\cite{RLMAC,RLMAC_multiagent}.

While AI is often executed in data centers, when used in large-scale networks (6G, Internet, \ldots), AI models need to be frequently (re-)trained and placed at many locations inside the network to meet, e.g., latency demands. We can no longer ignore where data, training, and inference are located but need to account for this explicitly: we need to build \emph{distributed AI applications}.  
Large-scale-network characteristics (i.e., latency, outages, convergence time after re-routing) will affect distributed AI applications much more than typical data-center networks, both during training and inference.


Therefore, distributed AI applications for large-scale networks need to be validated in realistic network environments that emulate or simulate these characteristics.
In this demo, we emulate networks with ContainerNet \cite{ContainerNet} and simulate aspects not amenable to emulation (e.g., wireless environments) with ns-3. 
We use this setup to demonstrate the testing of a distributed AI application in autonomic networks:
Hence, deep RL agents in the emulated network learn to design wireless MAC policies for a connected (simulated) wireless network. For resilient connectivity and discovery mechanisms, the emulated network uses the autonomic routing protocol KIRA~\cite{KIRA}.


\section{Testing Distributed AI in autonomic networks}

Our demo comprises three key components: support for distributed AI applications as such (Section~\ref{sec:testing_distr_AI}), KIRA as a connectivity/discovery solution for the control plane of autonomic networks (Section~\ref{sec:kira}), and AI for MAC learning as a concrete application (Section~\ref{sec:AI4MAC}).

\subsection{Testing Distributed AI Applications}
\label{sec:testing_distr_AI}

In distributed AI setups, 
training and inference must be analyzed separately due to their different requirements, e.g., regarding computing resources and latency. AI inference is usually less computationally intensive than AI training but can have tighter latency requirements. To reduce latency, AI inference could be deployed close to the point of use: 
distributed inference.

AI training requires a lot of data and computing resources. Network dynamics degrade AI model performance, requiring regular retraining. This is easily done locally, but to ensure the generalizability of a model, training data may need to come from multiple nodes. 
When training on a central node, the resulting traffic and data load from sending the training data has to be carefully weighed with the performance gains from AI. More likely, AI models train on or near data collectors in a distributed manner where model updates are exchanged between the nodes involved in the training, hopefully reducing data load and speeding up training. This so-called distributed training is, however, influenced by the network, e.g., through link failures or delays.

Therefore, when using AI models in a network, it is necessary to test its training and inference performance. Testing with actual networks is too expensive; simulations are often too abstract. 
Instead, we enhanced ContainerNet, a container-based network emulator, with support for GPU acceleration. This allows testing and validating AI applications in realistic network environments, enabling cost-efficient comparisons of different distributed AI use cases.


\subsection{KIRA: Resilient Connectivity and Discovery}
\label{sec:kira}
For distributed AI, nodes in a network
that collect data, train, or infer (e.g., routers, base stations, gateways, \ldots) need to be connected and know with whom they exchange data 
or model updates. 

In this demo, we deploy the autonomic routing protocol KIRA to provide scalable and resilient connectivity 
between components of distributed AI, i.e., between RL~agents and the (simulated) wireless environment. 
Furthermore, we use a distributed hash table (DHT) integrated into KIRA to provide an autonomic, distributed DNS service within the network.
In the demo, this is used as a simple discovery mechanism, e.g., to discover available RL~nodes.
These features allow the autonomic operation of distributed AI applications.


\subsection{AI for MAC Learning} 
\label{sec:AI4MAC}

The AI application in our demo
learns to design adaptive, environment-aware MAC policies in wireless networks, similar to \cite{RLMAC,RLMAC_multiagent}. 
This method allows dynamic changes to protocol features to meet various application requirements and optimize network performance.
The deep RL~agents learn typical MAC functions and policies through interaction with a simulated wireless network. By potentially adding, removing, or modifying protocol features, the protocol can continuously adapt and optimize based on the network and radio environment.




\section{Demo Setup}
\begin{figure*}[!ht]
    \centering
    \includegraphics[width=\linewidth]{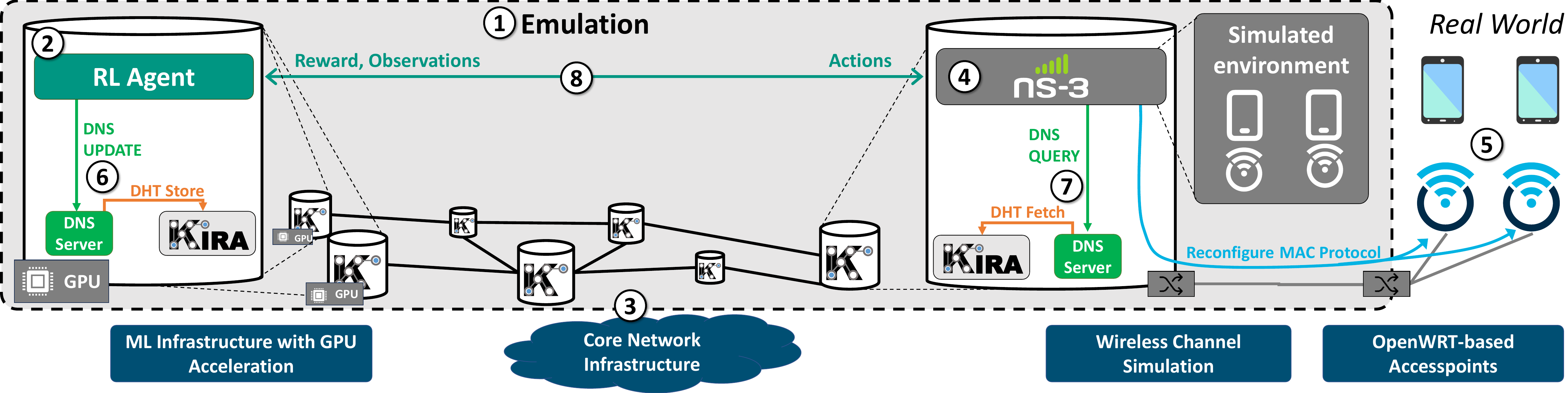}
    \caption{\centering Demo Setup}
    \label{fig:architecture}
\end{figure*}

Our demo setup is illustrated in \cref{fig:architecture}. It consists of a network emulated with ContainerNet\circled{1}, an integrated ns-3 simulation of the wireless network environment\circled{4}, and real hardware APs\circled{5}. 

In the emulated network itself, two RL~nodes\circled{2} access a GPU and are more suitable for training and inferring ML models. They are connected with a core network infrastructure consisting of KIRA routers\circled{3}.
This infrastructure also connects the RL~nodes to the node running the ns-3 simulation of the wireless environment\circled{4}. This node acts as gateway between the simulated wireless environment and the emulated network. Furthermore, this node has an OpenWRT interface to test the learned MAC protocol settings on actual hardware\circled{5}. 

All nodes of the emulated network run a KIRA routing daemon providing 1.) autonomic IPv6 connectivity between all nodes and 2.) a custom, local DNS server that uses KIRA's DHT to process DNS requests.
After short convergence, nodes send a DNS update\circled{6}, with a name (e.g., \emph{rlagent.kira.internal}) and their IPv6 address, to the local DNS server, which stores  
them in the DHT.
This allows the simulation node to discover the RL~nodes using a DNS query\circled{7}.
On the RL~nodes, deep RL~agents learn to design application-specific MAC protocols
by exchanging actions (MAC configurations), observations, and rewards\circled{8} with the wireless environment via the core network.
With each iteration, the simulated wireless environment changes, e.g., in terms of the number of access points (APs) and traffic. 
The same process takes place during inference, except that the policies of the deep RL~agents do not change. 

Using the demo setup, we demonstrate three distributed training and inference scenarios in a wireless environment with four APs. At the start-up of the network, we show how the nodes connect and discover each other in an autonomic manner via KIRA. 
We then compare the scenarios by evaluating the episode return of the deep RL~agents during training and the mean throughput of the wireless network during inference:  
\begin{itemize}
\item \emph{Central Single-Agent DRL}: A deep RL~agent trains and infers a policy on one RL~node to configure all APs.
\item \emph{Central Multi-Agent DRL}: A deep RL~agent trains and infers a policy on one RL~node for each AP individually.
\item \emph{Distributed Single-Agent DRL}: At each RL~node, a deep RL~agent receives observations and rewards for a subset of APs to train and infer a policy for its subset. 
\end{itemize}

\bibliographystyle{ieeetr}
\bibliography{literature}



\end{document}